\documentclass[prb,aps,twocolumn,superscriptaddress,showpacs,showkeys]{revtex4}

\usepackage{graphicx}
\usepackage{amsmath}
\usepackage{color}
\usepackage{amssymb}
\usepackage{verbatim}

\begin{document}

\title{ Remote Tomography Via von Neumann-Arthurs-Kelly Interaction}
\author{S. M. Roy}
\affiliation{ Homi Bhabha Centre for Science Education,\\
Tata Institute of Fundamental Research, Mumbai}
\author{Abhinav Deshpande}
\affiliation{Department of Physics, IIT Kanpur}
\author{Nitica Sakharwade}
\affiliation{Department of Physics, IIT Kanpur}

\date{\today}

\begin{abstract} 
Teleportation usually involves entangled particles 1,2  shared by Alice and Bob, Bell-state measurement 
on particle 1 and system particle by Alice, classical communication to Bob, and unitary 
transformation by Bob on particle 2. We propose a novel method: interaction-based 
remote tomography. Alice arranges an entanglement generating von Neumann-Arthurs-Kelly interaction 
between the system and two apparatus particles, and then teleports the latter to Bob. Bob reconstructs 
the unknown initial state of the system not received by him by quadrature measurements on the 
apparatus particles .



\end{abstract}

\maketitle


{\bf Introduction}. The idea of `quantum tracking' of a single system observable by an apparatus observable first 
occurred in the measurement theory of Von Neumann \cite{vonN},
and generalized to two canonically conjugate observables by Arthurs and Kelly Jr.\cite{AK}.
Suppose the initial state of the system-apparatus combine is factorized . If after interaction, the apparatus 
observable $X$ has the same expectation value in the final state  as the system 
observable $A$ in the initial state, for arbitrary initial state of the system, then 
$X$ is said to track $A$. This nomenclature was probably used first by Arthurs and Goodman\cite{AG} who , 
as well as, Gudder, Hagler, and Stulpe \cite{AG} proved the joint measurement uncertainty relation. 
The Arthurs-Kelly interaction  can also enable exact measurements of some quantum correlations 
between position and momentum \cite{SMR}.

We shall be  concerned here not with joint measurements but with the completely different idea of 
`remote quantum tomography' which is akin to `quantum teleportation' or the replication of an unknown quantum 
state of a particle at a distant location without physically transporting that particle. Teleportation, as  
first proposed by Bennett, Brassard, Cr$\acute{e}$peau, Jozsa, Peres and Wootters \cite{Bennett} and 
generalized to infinite dimensional Hilbert spaces by Vaidman \cite{Vaidman}, usually involves four different 
technologies.(i) An EPR-pair is shared by observers $A$ (Alice) and $B$ (Bob) at distant locations.
(ii) The system particle with unknown state is received by $A$ who makes a Bell-state measurement on the 
joint state of that particle and the first particle of the EPR-pair and (iii) communicates the result via a 
classical channel to $B$ , (iv) $B$ then makes a unitary transformation depending on the classical information 
on the second particle of the EPR-pair to replicate the unknown system state. Teleportation has been experimentally 
realized, e.g. by Bouwmeester et al\cite{Bouwmeester}, and the methods and uses extensively reviewed, e.g. by 
Braunstein et al\cite{Braunstein}. The density matrix of the system particle can be constructed by quadrature 
measurements on the second particle of the EPR pair completing remote tomography.

{\bf Interaction-based Remote Tomography}. 
We report here a completely new method for remote quantum tomography which replaces the above four 
technologies by the single step of an interaction between the system particle 
(say photon) and two apparatus photons. At location A, a system photon with unknown state  
interacts via a quantum optically generated Arthurs-Kelly interaction (see e.g. Stenholm \cite{AK}) 
with two apparatus particles (say photons) in a known state. The  apparatus photons are then 
sent to a distant observer $B$. $B$ makes quantum tomographic quadrature measurements on the apparatus photons and 
reconstructs the exact initial density matrix of the system photon without ever having received that particle.
(See Fig.1). Practical implementation 
will require a quantum channel to send the two apparatus photons from location $A$ to the distant location of $B$ and a   
generalization of single photon {\it Optical Homodyne Tomography } (see e.g. \cite{Vogel}, 
\cite{Braunstein-Leonhardt} and \cite{Yuen} ) to two photons , both of which seem feasible and worthwhile . 
Instead of the usual method of preparing the apparatus photons in an initial entangled state and sharing them 
between $A$ and $B$, this method of remote quantum tomography exploits the entanglement between the system 
photon and the apparatus photons generated by the three-particle Arthurs-Kelly interaction. Multiparticle 
interactions to generate entanglement have previously been exploited for  quantum enhanced 
metrology \cite{Roy-Braunstein}. We proceed now to put the new method on a rigorous footing. 

\begin{figure}[ht]
\begin{center}
 \includegraphics [width=\columnwidth]{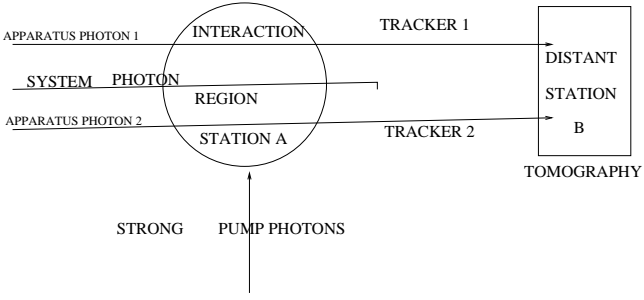}
\caption{Remote Quantum Tomography via Von Neumann-Arthurs-Kelly interaction between system photon and tracker photons. }
\end{center}
\end{figure}

 {\bf A Symmetry Property}. We shall use the Arthurs-Kelly system-apparatus interaction Hamiltonian ,
which is invariant under a class of simultaneous transformations on the system and apparatus specified below,
\begin{equation}
H=K (\hat{q}\hat{p}_1 +\hat{p}\hat{p}_2)= K (\hat{q}_{\theta }\hat{p}_{1,\theta} +\hat{p}_{\theta }\hat{p}_{2,\theta})
\end{equation}
where $K$ is a coupling constant , $\hat{q},\hat{p}$ are position and momentum operators of the system, 
$\hat{x}_1,\hat{x}_2$ are 
two commuting position operators of the apparatus (e.g. two photons), with conjugate momenta $\hat{p}_1,\hat{p}_2$ 
which are coupled to $\hat{q}$ and $\hat{p}$ respectively.The rotated quadrature operators with subscript $\theta$ 
are defined using the rotation matrix $R$,
\begin{equation}
 \begin{pmatrix}
 \hat{q}_{\theta }\\ 
\hat{p}_{\theta } 
\end{pmatrix}  =  R \begin{pmatrix} 
\hat{q}\\ 
\hat{p} 
\end{pmatrix},\>\begin{pmatrix}
\hat{p}_{1,\theta}\\ 
\hat{p}_{2,\theta}
\end{pmatrix}  =  R \begin{pmatrix} 
\hat{p}_1 \\ 
\hat{p}_2
\end{pmatrix},R= \begin{pmatrix}
 \cos \theta & \sin \theta \\
 -\sin \theta & \ cos \theta
\end{pmatrix}.
\end{equation}
The operators $\hat{p}_{j,\theta}$ are seen to be just the commuting 
momentum operators of the apparatus particles corresponding to rotated co-ordinates $x_{j,\theta}$, for $j=1,2$,
\begin{equation}
 x_{1,\theta}+i x_{2,\theta}= exp(-i\theta) (x_1 +i x_2),\hat{p}_{j,\theta}= -i \partial /\partial x_{j,\theta}.
\end{equation}
 
We also define,
\begin{equation}
\hat{x}_{1,\theta}+i \hat{x}_{2,\theta}=exp(-i\theta)( \hat{x}_1 + i \hat{x}_2).
\end{equation}

Then, in the case of the apparatus being two photons with annihilation operators $a_i$ ,$i=1,2$, 
\begin{equation}
 \hat{x}_{i,\theta}=a_i \exp{( -i\theta )}/\sqrt{2} +h.c. ,\> \hat{p}_{i,\theta}= \hat{x}_{i,\theta +\pi/2}.
\end{equation}

 {\bf Exact Solution of the Schr$\ddot{o}$dinger equation with generalized initial conditions}. We assume 
the constant $K$ to be so large that the free Hamiltonians of the system and the apparatus are negligible 
compared to $H$ during interaction time $T$.
We start from an initial factorized state , 
\begin{equation}
\langle q|\langle x_1,x_2|\psi (t=0)\rangle = \langle q| \phi \rangle  \chi (x_1,x_2) ,
\end{equation}
where $ \langle q| \phi \rangle $ is the unknown system wave fuction, and the apparatus wave function is
chosen to be a product of two Gaussians, $\chi (x_1,x_2)= \chi_1(x_1) \chi_2(x_2) $,
\begin{eqnarray}
 \chi_1(x_1)=\pi^{-1/4} b_1^{-1/2} \exp {[-x_1^2 /(2 b_1^2)]} \nonumber\\
 \chi_2(x_2)=\pi^{-1/4} (2 b_2)^{1/2} \exp {[-2 b_2^2 x_2^2 ]} .
\end{eqnarray}
 Arthurs and Kelly chose $b_2=b_1=b$. We solve the Schr$\ddot{o}$dinger equation with arbitrary  $b_1,b_2$ ;
 we need  $b_1\neq b_2$ to utilise the symmetry of the Hamiltonian.

 The commutator of the two terms in $H$ in fact commutes with each of the terms. Hence,
\begin{eqnarray}
\exp {(-iHt)}=\exp {(-iKt \hat{q}\hat{p}_1)} \nonumber\\\ 
\times exp {(-iKt \hat{p}\hat{p}_2)} \exp {(iK^2 t^2 \hat{p}_1 \hat{p}_2 /2)} \>.
\end{eqnarray}
If we work in the $q,x_1,p_2$ representation, the three exponentials on the right-hand side 
successively translate $x_1, q,x_1$ acting on the initial wavefunction. Hence the exact solution of 
the Schr$\ddot{o}$dinger equation is, 
\begin{eqnarray}
\langle q,x_1,p_2|t\rangle &=&  \chi _1 (x_1 -q Kt+ (1/2) p_2 K^2 t^2) \nonumber \\
& &\tilde \chi _2 (p_2) \phi (q-p_2 Kt), 
\end{eqnarray}
where $\tilde \chi _2 $ denotes a Fourier transform of $\chi _2$. The co-ordinate space wave function is 
given by a Fourier transform. Choosing $KT=1$ we obtain,
\begin{equation}
 \psi (q,x_1,x_2) = \int \psi (q,x_1,x_2,\xi) d\xi \>,
\end{equation}
 where,
\begin{eqnarray}
& & \psi (q,x_1,x_2,\xi) = \phi (\xi) \exp {(i(q-\xi)x_2) }/(2\pi \sqrt{b_1b_2}) \nonumber \\
& & \exp {(- \frac{(2x_1-q-\xi)^2}{8b_1^2} - \frac{(q-\xi)^2}{8b_2^2})}.
\end{eqnarray}
Tracing the system-apparatus density matrix over the system co-ordinate we obtain the apparatus density matrix 
at time T,
\begin{eqnarray}
& &\langle x_1,x_2|\rho _{APP}(T)|x_1'x_2'\rangle= \int \psi (q,x_1,x_2,\xi) \nonumber\\
& &\psi^{*} (q,x_1',x_2',\xi')dqd\xi d\xi '\>.
\end{eqnarray}
 The probability densities $P_1(x_1)$ and $P_2(x_2)$ for $x_1$ and $x_2$ are obtained by integrating 
the diagonal elements of this density operator over $x_2$ and $x_1$ respectively.In fact $P_1(x_1)$ and 
$P_2(x_2)$ can be obtained from the Arthurs-Kelly expressions by 
$b^2 \rightarrow (b_1 ^2 +b_2 ^2)/2$ and $b^{-2} \rightarrow (b_1 ^{-2} +b_2 ^{-2})/2$ respectively.The resulting 
expectation values of  $x_1,x_2$ equal those of $q,p$ respectively, but the dispersions are higher,
$ (\Delta x_1)^2=(\Delta q)^2 + (b_1 ^2 +b_2 ^2)/2,\> (\Delta x_2)^2=(\Delta p)^2 + (b_1 ^2 +b_2 ^2)/
(8b_1^2 b_2^2)$.

Our key new results require $b_1\neq b_2 $. First, integrating the off-diagonal elements of the 
apparatus density matrix over $x_2,x_2' $ ,
\begin{eqnarray}
& \int \langle x_1,x_2|\rho _{APP}(T)|x_1'x_2'\rangle dx_2 dx_2'= 
\frac{1}{b_1 b_2} \int |\phi (q)| ^2  \nonumber\\
&\exp {(- \frac{(x_1-q)^2 +(x_1'-q)^2 }{2b_1^2} )} dq \>.
\end{eqnarray}
This shows that we can extract the exact initial system position probability density from 
the final apparatus density matrix as the expectation value of an apparatus observable. 
\begin{eqnarray}
\Bigl \lvert \langle q=x_1\vert \phi \rangle \Bigr \rvert ^2= lim_{b_1 \rightarrow 0 } 
\frac{b_2}{\sqrt{\pi}}\int dx_2 dx_2' \nonumber \\
\langle x_1,x_2|\rho _{APP}(T)|x_1 x_2'\rangle \nonumber \\
=lim_{b_1 \rightarrow 0 }  Tr \rho _{APP}(T) Y(x_1),
\end{eqnarray}
where $Y(x_1) $ is the apparatus observable,
\begin{eqnarray}
Y(x_1)= \frac{b_2}{\sqrt{\pi}} |x_1\rangle \langle x_1| 
\int |x_2'\rangle \langle x_2''|dx_2'dx_2'' \nonumber \\
=2b_2\sqrt{\pi} (|x_1\rangle \langle x_1|)(|\hat{p}_2=0\rangle \langle \hat{p}_2=0|).
\end{eqnarray}
Similarly, the exact initial 
system momentum probability density is an expectation value of an apparatus observable
in the final apparatus density matrix,
 \begin{eqnarray}
 \Bigl \lvert \langle p=x_2\vert \phi \rangle \Bigr \rvert ^2 = lim_{b_2 \rightarrow \infty } 
\frac{1}{2b_1\sqrt{\pi}}\int dx_1 dx_1' \nonumber \\
\langle x_1,x_2|\rho _{APP}(T)|x_1' x_2\rangle \nonumber \\
=lim_{b_2 \rightarrow \infty }  Tr \rho _{APP}(T) Z(x_2),
\end{eqnarray}
where $Z(x_2) $ is the apparatus observable,
\begin{equation}
Z(x_2)= \frac{\sqrt{\pi}}{b_1} (|x_2\rangle \langle x_2 |)(|\hat{p}_1=0\rangle \langle \hat{p}_1=0|).
\end{equation}  
In the limit, $ b_1 \rightarrow 0,\>b_2 \rightarrow \infty$,
we have faithful tracking of both system position and system momentum, since $Y(x_1)$  
tracks the position projectors $|\hat{q}=x_1><\hat{q}=x_1|$ for all $x_1 $ and $Z(x_2)$ tracks the system 
momentum projectors  $|\hat{p}=x_2><\hat{p}=x_2|$ for all $x_2 $. 

Further, the Wigner function of the initial system state can be calculated exactly from the 
final apparatus density matrix,
\begin{eqnarray}
 W(x_1,x_2)= lim _{b_1 \rightarrow 0 ,b_2 \rightarrow \infty } \frac{b_2}{2\pi b_1} \nonumber \\ 
\int dx_1' dx_2'\langle x_1,x_2|\rho _{APP}(T)|x_1' x_2'\rangle .
\end{eqnarray}
We now show that we can indeed measure a continuous infinity of apparatus observables on the final state 
to obtain the initial Wigner function of the system particle. 
 
{\bf Rotated quadratures and Quantum Tomography}.
In order to harness the symmetry property mentioned above,
we need a corresponding symmetry property of the initial apparatus state,
$\chi (x_1,x_2) =\chi (x_{1,\theta},x_{2,\theta})$.
Therefore we are forced to use  initial apparatus states very different from Arthurs and Kelly. We need,
\begin{eqnarray}
2 b_1 b_2 =1;\> \chi (x_1,x_2) =\chi (x_{1,\theta},x_{2,\theta})\nonumber\\
= \pi^{-1/2} b_1^{-1} \exp {[-(x_1^2 + x_2^2)/(2 b_1^2)]}.
\end{eqnarray}
For this choice , the system-apparatus initial state can be rewritten for arbitrary $\theta$ as, 
\begin{eqnarray}
\langle \hat{q}_{\theta }=q_{\theta }|\langle \hat{x}_{1,\theta}= x_{1,\theta},\hat{x}_{2,\theta}=x_{2,\theta}
|\psi (t=0)\rangle \nonumber\\
= \langle \hat{q}_{\theta }=q_{\theta }|\> \phi \rangle   \chi (x_{1,\theta},x_{2,\theta}) ,\>
\end{eqnarray} 
with the obvious notation $(\hat{q}_{\theta }-q_{\theta }) |\hat{q}_{\theta }=q_{\theta }\rangle =0 $.
Since the Hamiltonian $H$ and the initial apparatus states have exactly the same form in terms 
of the rotated variables as in terms of the original variables,
we can repeat the previous calculations with $\hat{q}_{\theta }, 
\hat{p}_{\theta }, q_{\theta },p_{\theta },x_{1,\theta},x_{2,\theta}$ replacing 
$\hat{q}, \hat{p} ,q,p,x_1,x_2$ respectively. Hence the matrix elements of $\rho_{APP.} $ are obtained by 
replacing in the previously obtained expressions 
$$ q,p,x_1,x_2,x'_1,x'_2 \rightarrow q_\theta,p_\theta,x_{1,\theta},x_{2,\theta},x'_{1,\theta},x'_{2,\theta}. $$
Thus, we obtain for arbitrary $\theta$ ,
\begin{equation}
\Bigl \lvert \langle \hat{q}_\theta=u\vert \phi \rangle \Bigr \rvert ^2 = 
lim_{b_1 \rightarrow 0 }  Tr \rho _{APP}(T) Y_{\theta}(u),
\end{equation}
\begin{eqnarray}
 Y_{\theta}(u)\equiv \frac{\sqrt{\pi}}{b_1} |\hat{x}_{1,\theta}=u\rangle
 \langle \hat{x}_{1,\theta}=u| \nonumber\\
|\hat{p}_{2,\theta}=0\rangle \langle \hat{p}_{2,\theta}=0|.
\end{eqnarray}
Since, $ \hat{p}_\theta = \hat{q}_{\theta+\pi/2}$ 
the initial system probability densities for it are obtained from above just by replacing 
$ \theta \rightarrow  \theta +\pi/2$.
  
We have proved that in the limit,
\begin{equation}
  b_1 \rightarrow 0 , b_2 =1/(2b_1)\rightarrow \infty ,
\end{equation}
we can recover exactly the initial system probability densities of arbitrary Hermitian linear
combinations $\hat{q}_{\theta }$,
\begin{equation}
 \langle \hat{q}_{\theta}=u|\> \rho_S |\hat{q}_{\theta}=u\rangle =
\Bigl \lvert \langle \hat{q}_\theta=u\vert \phi \rangle \Bigr \rvert ^2
\end{equation}
and hence the initial Wigner function, by measuring expectation values of Hermitian operators in the same 
final state of the apparatus after interaction. 

{\bf Reconstruction of the initial Density Matrix of the System from the final Apparatus Density Matrix}.
Quantum tomography is 
completed by calculating the Wigner function $W(q,p)$ as an inverse Radon transform, 
\begin{eqnarray}
 W(q,p)=(2\pi)^{-2}\int_0 ^\infty \eta d\eta \int_0 ^{2\pi}d\theta \int_{-\infty}^{\infty} du \nonumber \\
\exp{(i\eta (u-(q\cos {\theta} + p\sin{\theta } ) ) )} \langle \hat{q}_{\theta}=u|\> \rho_S |\hat{q}_{\theta}=u\rangle,
\end{eqnarray}
and from that the density operator, 
\begin{eqnarray}
 \langle q |\rho_S |q'\rangle =(2\pi)^{-1} \int_0 ^{\pi} |q-q'| d\theta (\sin{\theta})^{-2} \nonumber\\
 \exp {((-i(q^2-q'^2 )\cot{\theta} )/2)} \int_{-\infty}^{\infty} du \nonumber\\
\exp{(iu(q-q')/\sin{\theta } )} \langle \hat{q}_{\theta}=u|\> \rho_S |\hat{q}_{\theta}=u\rangle .
\end{eqnarray}

{\bf Accounting for time evolution of the apparatus photons during transit time $\tau$ to distant location B}.
Note that 
\begin{eqnarray}
 Tr \rho _{APP}(T) Y_{\theta}(u)=Tr \rho _{APP}(T+\tau) \nonumber\\ 
\times exp(-iH_0 \tau)Y_{\theta}(u)exp(iH_0 \tau),
\end{eqnarray}
where the Hamiltonian $H_0=\omega (a_1\dagger a_1 +a_2\dagger a_2 +1) $, if  the photons have the same frequency 
$\omega$. Hence the $ \langle \hat{q}_{\theta}=u|\> \rho_S |\hat{q}_{\theta}=u\rangle $ are equivalently given by 
replacing 
\begin{eqnarray}
 \rho _{APP}(T), \hat{x}_{1,\theta},\hat{p}_{2,\theta}\>\rightarrow \>\rho _{APP}(T+\tau),\nonumber \\
cos(\omega \tau) \hat{x}_{1,\theta} -sin(\omega \tau) \hat{p}_{1,\theta},
cos(\omega \tau)\hat{p}_{2,\theta}+ sin(\omega \tau)\hat{x}_{2,\theta}\nonumber
\end{eqnarray}
 respectively.We just have to measure 
different quadratures for the apparatus photons depending on the transit time $\tau$.

 \begin{figure}[thb]
\begin{center}

\includegraphics [width=\columnwidth]{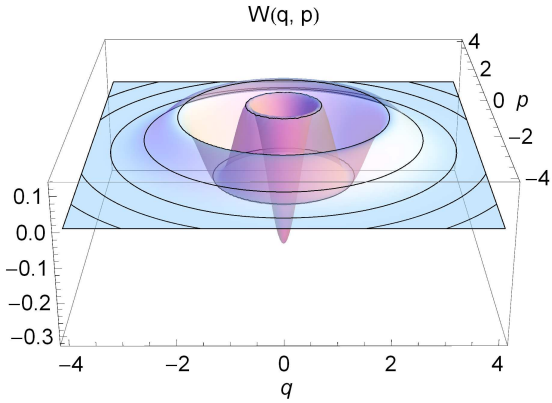}
\caption{The Wigner function for the 3$ ^{rd} $ excited state of the harmonic oscillator}

\end{center}
\end{figure}

\begin{figure}[thb]
\begin{center}
\includegraphics [width=\columnwidth]{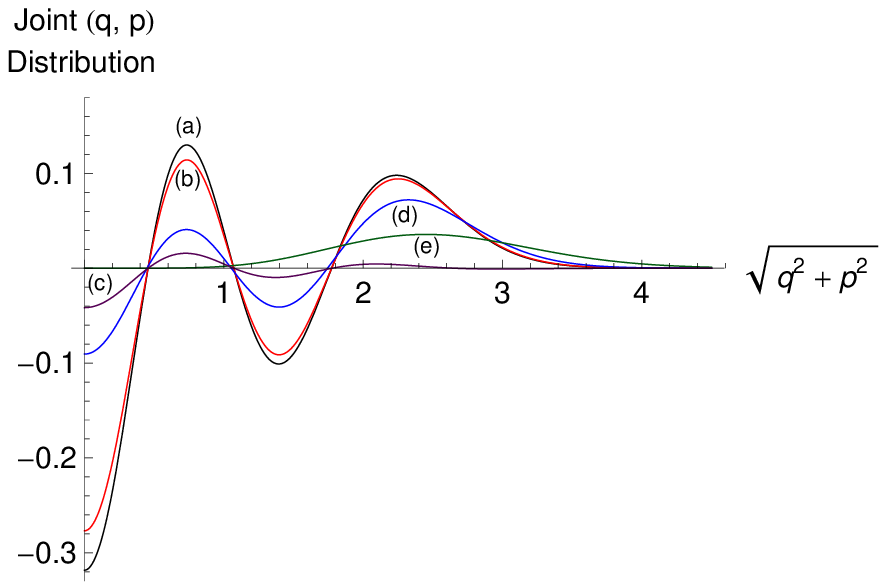}
\caption{Joint distributions in ($ q,p $) for the third excited state of the oscillator 
as a function of $ \sqrt{q^2+p^2} $ 
(a): Wigner function (b): Reconstructed Wigner function with $ b_1 = 0.1$. 
(c): Difference between curves (a) and (b). (d): Reconstructed Wigner function with $ b_1 = 0.3$. 
(e): Arthurs-Kelly probability distribution .}

\includegraphics [width=\columnwidth]{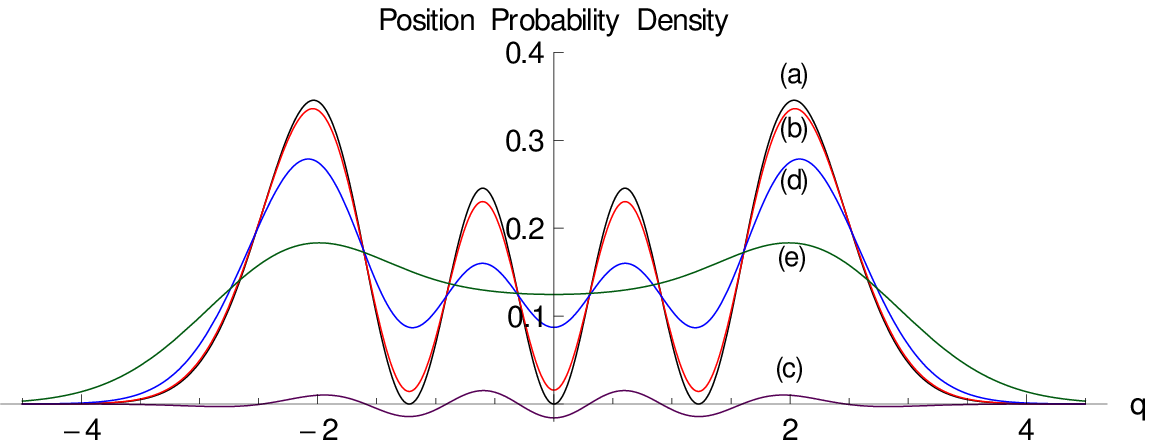}
\caption{Position probability densities in for the third excited state. (a): Quantum probability density of the state. 
(b): Obtained from reconstructed Wigner function with $ b_1 = 0.1$. (c): Difference between curves (a) and (b). 
(d): Obtained from reconstructed Wigner function with $ b_1 = 0.3$. 
(e): Obtained from Arthurs-Kelly probability distribution .}

\end{center}
\end{figure}

{\bf Quantitative comparisons for the third excited state of the oscillator .}

Our exact theorems are for the limit $b_1\rightarrow 0$. The purpose here is to estimate how small this 
parameter has to be for reasonably accurate reconstruction of the initial state which ,in this example, 
is chosen to be the highly non-classical third excited of the oscillator. 
The wave function in  the position basis is
\begin{equation}
\phi(q) = (2q^3-3q)exp\left(-\frac{q^2}{2}\right)/(\sqrt{3}\ \pi^{1/4}).
\end{equation}
The Wigner function is a function of $q^2+p^2 \equiv d$
\begin{equation}
W(d) = exp(-d)[4d^3-18d^2+18d-3]/(3\pi).
\end{equation}




In the figure we make quantitative comparisons between the Wigner function, our reconstructed Wigner function
with $ 2b_1 b_2 = 1$ (for $ b_1 = \{0.1, 0.3\}$) and the Arthurs-Kelly 
Probability distribution .It is worth noting 
that for $ b_1 = \frac{1}{\sqrt{2}} $, the reconstructed Wigner function is equal to 
the Arthurs-Kelly distribution which differs greatly from the true Wigner function. Towards practical utility, 
note that  for $b_1=.1$ the reconstructed Wigner function and the position probability derived from it  
are already very close to the actual, though the theorem of exact equality is only in 
the limit $b_1 \rightarrow 0$.  

  
{\bf Conclusions and Outlook}. (i) We have shown that the generation of entanglement by the Arthurs-Kelly Hamiltonian 
between an unknown state of a system photon and chosen initial state of two apparatus photons enables 
a one-step remote tomographic reconstruction of the unknown initial state of the system photon , instead of the usual 
four step process. This  `interaction based remote tomography' is practically feasible because the 
technology of generating this interaction quantum optically is well established. 

(ii) Remote Tomography requires the measurement of the two photon observable  $Y_{\theta}(u) $ . Since 
this is a product of two commuting quadrature operators for the apparatus photons, each of the kind usually 
measured for a single photon, the measurement should be possible by generalizing optical homodyning 
to the two teleported photons. This generalization will by itself be a stimulating development. 

(iii) The Arthurs-Goodman result on impossibility of simultaneous accurate 
tracking of position and momentum by commuting observables of the apparatus is not violated. 
The secret is that the apparatus observables tracking position and momentum do not commute,
$$
[Y(x_1), Z(x_2)]  \neq 0 .
$$
This is not a problem since we are only interested in faithful tomography of the initial system state , 
from repeated measurements on the teleported apparatus particles, 
and not in the simultaneous measurement of position and momentum.

(iv) The final density operator of the system can also be exactly calculated and it can be seen that 
$<q>_T =<q>_0$ , $\Delta q^2 _T =\Delta q^2 _0 + 2 b_2 ^2$; since the final system state is different from 
the initial state, and depends on the initial states of both the system and the apparatus, the no-cloning 
\cite{Wootters} and no-hiding theorems \cite{Braunstein-Pati} are respected. 
 
(v) If the initial system $S_1$ is entangled with another system $S_2$, the apparatus photons after interaction 
with $S_1$ become entangled with $S_2$, leading to interaction-based teleportation of entanglement \cite{SMR1}.
 
{\bf Acknowledgements}. SMR thanks Sam Braunstein for many helpful suggestions including the name 
'remote tomography', and Arun Pati, Ujjwal Sen and Aditi Sen De for discussions. AD and NS thank 
the NIUS program of the Homi Bhabha Centre for Science Education ; SMR thanks the Indian National 
Science Academy for the INSA Senior Scientist award.

\end{document}